# Title

## Origin of anomalous p-type conductivity in monolayer Fe-doped MoS$_2$

## Authors

Xiangning Quan[1,†], Xiaoqiu Yuan[2,†], Junwei Zhang[1,†], Xuebing Peng[3,†], Helin Mei[3,†], Cheng Yan[4], Hong Zhang[5,6], Hongli Li[7], Daqiang Gao[3], Yongjian Wang[1], Mingsu Si[1], Lili Zhang[1,*], Anmin Zhang[3,*], Zongyuan Zhang[2,*], Lei Shan[2], and Yong Peng[1,*]

## Affiliations

[1]School of Materials and Energy, or Electron Microscopy Centre of Lanzhou University, Lanzhou University, Lanzhou, China.

[2]Information Materials and Intelligent Sensing Laboratory of Anhui Province, Center of Free Electron Laser and High Magnetic Field, Institutes of Physical Science and Information Technology, Anhui University, Hefei, China.

[3]School of Physical Science and Technology, Lanzhou University, Lanzhou, China.

[4]School of Chemistry, Faculty of Science, The University of Sydney, Sydney, Australia.

[5]Yunnan Key Laboratory of Electromagnetic Materials and Devices, School of Materials and Energy, Yunnan University, Kunming, China.

[6]Electron Microscopy Center, Yunnan University, Kunming, China.

[7]State Key Laboratory of Solid Lubrication, Lanzhou Institute of Chemical Physics, Chinese Academy of Sciences, Lanzhou, China

*Corresponding author. Email: zll@lzu.edu.cn (L.Z); amzhang@lzu.edu.cn (A.Z.); zongyuanzhang@ahu.edu.cn(Z.Z.); pengy@lzu.edu.cn (Y.P.)

†These authors contributed equally.

**Abstract**

Substitutional doping effectively modulates carrier polarity of semiconducting two-dimensional (2D) transition metal dichalcogenides (TMDs) like $MoS_2$. Although Fe doping typically induces n-type conductivity in monolayer $MoS_2$, anomalous p-type behavior has also been experimentally reported, the origin of which remains unresolved. Here, we prove that this anomalous p-type conductivity originates from defect associates formed through interactions between Fe dopants and S atoms, which consists of three Fe substituting Mo ($Fe_{Mo}$) point defects arranged into an equilateral triangle with a central S atom, denoted as $3Fe_{Mo}$-S associate. Its p-type effect is directly verified through scanning tunneling microscopy/scanning tunneling spectroscopy (STM/STS) measurement, in sharp contrast to the n-type behavior induced by isolated $Fe_{Mo}$ point defects, and the conclusion is further supported by electrical transport measurements and first-principles calculations. Similar $3Fe_W$-S associates and their p-type doping effect are also identified in monolayer Fe-doped $WS_2$. This work resolves a longstanding controversy and highlights the critical role of defect associates in modulating properties of 2D TMDs.

**Teaser**

This work clarifies that the anomalous p-type conductivity induced by Fe doping in $MoS_2$ arises from the associates composed of Fe dopants and S atoms.

**MAIN TEXT**

**Introduction**

Semiconducting 2D TMDs, such as $MoS_2$ and $WS_2$, represent promising channel materials for post-Moore electronics due to their exceptional electrical properties arising from atomic thickness (*1-4*). Carrier polarity manipulation for 2D TMDs is significant and a prerequisite for construction of 2D p-n junctions to promote their real-world utilization. Defects play a critical, even decisive role in modulating 2D TMDs' electrical properties. So far, various types of point defects including vacancies (*5, 6*), substitutional dopants (*7-9*), antisite defects (*10, 11*), and interstitial atoms (*12, 13*) have been intensively investigated to regulate the carrier polarity of 2D TMDs, among which substitution doping is believed to be one of the most important and effective strategies for its high controllability and stability.

Transition metals serve as promising dopants for 2D TMDs because of their ability to form stable covalent bonds with the host lattice. Generally speaking, when transition metals such as Re and Mn (*14, 15*) with more valence electrons than that of transition metal M in $MX_2$ (M=Mo, W, X=S, Se) substitute M atoms, system will be lead to n-type doping (electron doping), meanwhile for transition metals with less valence electrons like V, Nb, and Ta, p-type doping (hole doping) is achieved (*16-18*). However, as a typical transition metal, Fe doping in $MoS_2$ brings contradictory results, in which both enhanced n-type doping and p-type doping has been observed and reported by different groups (*19-23*). It is widely accepted that isolated Fe dopants can enhance the n-type conductivity of $MoS_2$ because Fe has two more valence electrons than Mo and they completely or partially release these electrons when substitute Mo atoms. However, the origin of anomalous p-type doping is still unclear.

In this study, we provide direct evidence that the anomalous p-type doping in monolayer Fe-doped $MoS_2$ arises from the Fe dopant-related defect associates, which formed through interactions between Fe atoms and S atoms. The atomic-resolution high angle annular dark field scanning transmission electron microscopy (HAADF-STEM) images confirm the real-space atomic structure of the single associate is three nearest-neighboring $Fe_{Mo}$ point defects arranging into a triangular configuration with a S atom at the center, which is labelled as $3Fe_{Mo}$-S associate here. Stark contrast to isolated $Fe_{Mo}$ point defects that contribute electron doping to monolayer $MoS_2$ (*19-21*), the STM/STS measurements demonstrate that single $3Fe_{Mo}$-S associate induces acceptor states above the valence band maximum (VBM) of $MoS_2$, confirming its hole doping effect. This hole doping role is further verified by both electrical transport measurements and theoretical calculations. Similar $3Fe_W$-S associates and their hole doping effect are also observed in monolayer Fe-doped $WS_2$. This study reconciles a long-time inconsistency observed in Fe-doped $MoS_2$ and demonstrates the significance of defect associates in tailoring properties of 2D TMDs.

## Results

### Recognition of $3Fe_{Mo}$-S associates in monolayer Fe-doped $MoS_2$

Fe-doped and pristine $MoS_2$ monolayers were synthesized by chemical vapor deposition (CVD) method using $Fe_2Mo_3O_{12}$ and $MoO_3$ as precursors respectively (Methods and fig. S1 for details). Fe-doped $MoS_2$ monolayers exhibit a triangular morphology as displayed in optical image (inset, Fig. 1A). The height of a freshly prepared monolayer Fe-doped $MoS_2$ flake is determined to be 0.96 nm as measured by atomic force microscope (fig. S2), confirming its monolayer feature. Figure 1A shows the Raman spectra of monolayer Fe-doped and pristine $MoS_2$. Monolayer pristine $MoS_2$ shows two characteristic Raman peaks locating at 385.1 $cm^{-1}$ and 404.5 $cm^{-1}$, corresponding to the $E'$ and $A_1'$ modes, respectively (*24, 25*). After incorporating Fe atoms into $MoS_2$, two new peaks locating at 349.4 $cm^{-1}$ and 375.8 $cm^{-1}$ are observed, which come from zone-edge phonon modes induced by Fe doping-related disorders (Supplementary Text S1, fig. S3 and Table S1). Figure 1B shows the fine-scanning Fe 2p, Mo 3d, and S 2p X-ray photoelectron spectroscopy (XPS) spectra of Fe-doped (upper part) and pristine (lower part) $MoS_2$. The observation of Fe $2p_{1/2}$ peak (720.8 eV), Fe $2p_{3/2}$ peak (707.0 eV) along with its satellite peak (711.4 eV) in Fe-doped $MoS_2$ unambiguously confirm the Fe incorporation in monolayer $MoS_2$. Compared with XPS spectra of the pristine $MoS_2$, the peak positions of both Mo and S in Fe-doped $MoS_2$ are shifted 0.8 eV toward lower binding energies. This indicates that the introduction of Fe weakens Mo-S bonds and implies a downshift of the Fermi level ($E_F$) toward the VBM, which is a characteristic of hole doping for monolayer $MoS_2$ (*26*). In addition, the photoluminescence spectra, XPS survey scan, and energy dispersive X-ray spectrum (EDX) of samples collectively validate the incorporation of Fe into $MoS_2$ (figs. S4 to S6).

To confirm the accurate position of Fe atoms in monolayer $MoS_2$, we performed atomic-resolution imaging on the samples using spherical aberration-corrected scanning transmission electron microscopy (Cs-STEM). Figure 1C shows the enlarged atomic-resolution HADDF-STEM image of monolayer Fe-doped $MoS_2$ cut from the white dashed box area in another HAADF-STEM image (fig. S7D). Since the contrast intensity in HAADF-STEM image is proportional to the atomic number $Z^{1.7}$ (*27*), the brightest atoms in Fig. 1C can be indexed to Mo atoms. Mo atoms and 2S columns form the hexagonal crystal structure of monolayer $MoS_2$ as indicated in the upper left of Fig. 1C. Due to the smaller atomic number of Fe than Mo, its contrast intensity is lower than Mo and some darker spots in the sublattice of Mo sites are observed as highlighted with white circles in Fig. 1C, which correspond to Fe atoms. In order to verify this further, a simulated HAADF-STEM image by QSTEM software (*28*) containing an isolated $Fe_{Mo}$ point defect is presented in Fig. 1D. The experimental image, which is cut from the yellow rectangle in fig. S7D, corresponding to the simulated one is displayed in Fig. 1E. The intensity profile of five atoms along the yellow solid line in Fig. 1E (left to right: Mo-Mo-darker spot-Mo-Mo) closely matches with the red broken line in Fig. 1D (left to right: Mo-Mo-Fe-Mo-Mo) as displayed in Fig. 1F, confirming that the darker spot in Fig. 1E is Fe atom. This demonstrates that Fe atoms substitute Mo atoms and this substitution is achieved by a random way (fig. S7D). Besides isolated $Fe_{Mo}$ point defects, three nearest-neighboring Fe atoms arrange into a triangle can be found as marked with a yellow triangle in Fig. 1C. It is noted that the brightness of the center of the triangle corresponding to the sublattice of S atoms is approximately half that of the other equivalent sites, which may be caused by the loss of a single S atom from the central S-column site. To confirm this, similarly, a simulated HAADF-STEM image containing one triangle configuration is presented in Fig. 1G. Its corresponding experimental image cutting from the blue rectangle in fig. S7D is shown in Fig. 1H. The contrast intensity curve of five atomic positions measured along the light blue solid line in Fig. 1H is highly consistent with the contrast intensity curve measured along the dark blue broken line in Fig. 1G (left to right: 2S-2S-S-2S-2S) as shown in Fig. 1I, unambiguously confirming that only one S atom left in the center of the triangle. Thus, the atomic structure of this defect associate is determined, which is composed of three nearest-neighboring $Fe_{Mo}$ point defects and one S atom and is marked as $3Fe_{Mo}$-S associate. Its corresponding top view and side view of atomic configurations are displayed in Fig. 1J. This schematic diagram of atomic structure shows the lattice distortion of monolayer $MoS_2$ caused by Fe dopants. It is noticed that the S and Fe atoms in this structure tend to be coplanar due to the lowest system energy (Supplementary Text S2, fig. S10). Above analyses reveal that there are two defect types in monolayer Fe-doped MoS$_2$, one is isolated Fe$_{Mo}$ point defect and the other is 3Fe$_{Mo}$-S associate.

More isolated Fe$_{Mo}$ point defects and 3Fe$_{Mo}$-S associates in monolayer Fe-doped MoS$_2$ can be observed in fig. S7D, and the Fe doping concentration is calculated to be 6.9 at. % (Supplementary Text S3), in which Fe atoms forming 3Fe$_{Mo}$-S associates account for approaching 40% of the total Fe atoms (fig. S8). Moreover, monolayer MoS$_2$ with tunable Fe doping concentration can be synthesized by replacing Fe$_2$Mo$_3$O$_{12}$ precursor with FeS$_2$ and MoO$_3$ and by tuning their atomic ratio (fig. S9). Further analyses confirm that with the increase of FeS$_2$ content, both the Fe doping concentration and the proportion of Fe atoms in 3Fe$_{Mo}$-S associates gradually increase, providing possibility for the realization of defect concentration-dependent properties for MoS$_2$. In addition, the 3Fe$_{Mo}$-S associates in individual nanosheets exhibit predominant unidirectional alignment (figs. S7E and S9) as the central S atom effectively lowers their formation energy (Supplementary Text S4, figs. S11 and S12).

**Detection of the hole doping role of 3Fe$_{Mo}$-S associates for monolayer MoS$_2$**

After transferring monolayer Fe-doped MoS$_2$ flakes onto Au substrate and cleaning its surface by annealing under ultrahigh vacuum conditions at ~200 °C, we conducted the STM/STS characterizations in order to probe defect induced electronic states in the sample. Figure 2A shows the STM topography measured in the constant current mode. Stacking monolayer Fe-doped MoS$_2$ on an Au (111) substrate forms hexagonal moiré patterns as depicted in Fig. 2A, and the fast Fourier transform (FFT) in the inset confirms their six-fold symmetry. The observed moiré period is ~3.3 nm, which is consistent with previous studies on stacking monolayer MoS$_2$ on Au (111) (*29, 30*). The diagonal line in Fig. 2A comes from a substrate step on Au surface, which can be clearly seen in its corresponding 3D image (fig. S13A). Before STS measurements, identification of Fe$_{Mo}$ point defects and 3Fe$_{Mo}$-S associates in STM topography image is essential. Firstly, the simulated STM images for single Fe dopant in monolayer MoS$_2$ in a previous research showed that a triangular bright spot would appear near an isolated $Fe_{Mo}$ point defect in constant current mode, which comes from the local density of states of S atoms near the Fe atom (*31*). Based on this, the triangular bright spots in Fig. 2A can be indexed to $Fe_{Mo}$ point defects, and three isolated $Fe_{Mo}$ point defects in Fig. 2A marked with red circles are labelled as defect #1(1), defect #1(2) and defect #1(3), respectively. The distance between adjacent vertices of the bright triangle is ~1.2 nm (inset, Fig. 2C), which is twice as long as the distance between two adjacent vertices at the S vacancy (~0.7 nm) that has been observed (*32*), completely excluding the interference of S vacancies as the defect of this morphology. Secondly, sulfur vacancy-related defects typically appear as dark triangles in STM images (*33, 34*), coupling with the electronic state discussed later, we can identify the dark spot marked with defect #2(1) in Fig. 2A as a $3Fe_{Mo}$-S associate.

The differential conductance (dI/dV) spectra measured on top region (orange point) and hollow region (purple point) on monolayer pristine $MoS_2$ are presented in Fig. 2B. A gap ranging from -1.25 V to 0.25 V in the spectra of density of states (DOS) can be seen, which is the bandgap of monolayer pristine $MoS_2$ stacking on Au (111). The measured bandgap of 1.5 eV is smaller than that of the free-standing monolayer $MoS_2$ (~1.7 eV) for the strong hybridization between the electronic states of $MoS_2$ and Au substrate (*35*). The peak labelled as $\Gamma_1$ located at ~0.9 V is interpreted as a hybrid Au-$MoS_2$ or an interface state. The other peak $\Gamma_2$ located at ~1.4 V corresponds to the prominent resonances at the $\Gamma$ point in Brillouin zone, similar to the case in free-standing $MoS_2$. In the valence band, the peak measured at top spot region is shifted to higher voltage than that at hollow spot, which can be explained by the unequal hybridization between these states near Q point in Brillouin zone with Au substrate (*36*). The dI/dV spectra collected along the arrowed line (left inset, Fig. 2C) approaching the center of defect #1(1) are displayed in Fig. 2C. At position corresponding to the end of the arrow, that is, away from the center of defect #1(1), a DOS peak at 0.95 V still can be observed. With the approaching of the center of defect #1(1), this peak becomes stronger and shifts to lower voltage, which can be due to the bandgap decrease accompanied by an electron doping effect induced by isolated $Fe_{Mo}$ point defect (*34*).

The spectra of a 3Fe$_{Mo}$-S associate shown in Fig. 2D were also studied by the same way as that of Fe$_{Mo}$ point defect. As the center of defect #2(1) is approached, a DOS peak above the VBM of monolayer MoS$_2$ gradually appears at -0.8 V. The appearance of this peak can be contributed to the acceptor states induced by defect #2(1), drawing the conclusion that 3Fe$_{Mo}$-S associates introduce holes into the system. We also noticed other evolutions in the electronic states near defect #2(1) such as the increase in DOS at valence band edge around -1.5 V, which may originate from the upward shift of the valence band or the influence of moiré patterns as described in Fig. 2A. More STS measurements on other isolated Fe$_{Mo}$ point defects and 3Fe$_{Mo}$-S associates show similar results (fig. S13, B and C) with that in Fig. 2 (C and D). All above STS results demonstrate that the isolated Fe$_{Mo}$ point defect and 3Fe$_{Mo}$-S associate indeed produces electron doping and hole doping for monolayer MoS$_2$, respectively.

**Electrical transport properties of Fe-doped and pristine MoS$_2$ monolayers**

In order to verify above inference, the electrical transport properties of Fe-doped and pristine MoS$_2$ monolayers were measured. Figure 3A shows the schematic field-effect transistor (FET) device structure made of monolayer Fe-doped MoS$_2$. Figure 3B presents the output characteristics of monolayer Fe-doped MoS$_2$, along with an optical image of a representative device (inset). The Fe-doped monolayer exhibits highly linear output behavior, in contrast to the pronounced nonlinearity observed in pristine MoS$_2$ (fig. S14), indicating that Fe doping significantly improves the electrical contact between the semiconductor and the metal electrodes. Figure 3 (C and D) shows the transfer curves of monolayer pristine and Fe-doped MoS$_2$, respectively. The pristine sample exhibits well n-type semiconductor behavior with an on/off ratio of ~10$^6$ at V$_{ds}$=5.0 V, while the Fe-doped sample shows degenerate p-type conductivity behavior. Contrary to the established view that Fe$_{Mo}$ point defects act as electron donors in monolayer MoS$_2$ for more valence electrons (Fe: [Ar]3d$^6$4s$^2$, Mo: [Kr]4d$^5$5s$^1$), our Fe-doped samples exhibit degenerate p-type conductivity. We attribute this to the compensatory hole doping from 3Fe$_{Mo}$-S defect associates, which outweighs the electron donation from isolated Fe$_{Mo}$ point defects. This interpretation is consistent with above STS results.

**Exploration on the origin of hole doping of 3Fe$_{Mo}$-S associate**

To understand the n-type and p-type conductivities induced by different defect configurations in monolayer Fe-doped MoS$_2$, we performed first-principles calculations of the electronic band structures of pristine monolayer MoS$_2$ as well as systems containing an isolated Fe$_{Mo}$ point defect or a 3Fe$_{Mo}$-S associate. A 4×4×1 supercell of monolayer MoS$_2$ was chosen to mimic the desired defects (fig. S11, A and B). The pristine monolayer MoS$_2$ exhibits a direct bandgap of 1.71 eV (fig. S15), consistent with previous reports (*37*). Figure 4A shows the band structure of monolayer MoS$_2$ containing one isolated Fe$_{Mo}$ point defect, demonstrating that substituting one Mo atom with Fe introduces four flat defect bands within the band gap near the $E_F$. Our calculated results found that the metal-S bond length is reduced from 2.41 Å (Mo-S bond) in pristine MoS$_2$ to 2.28 Å (Fe-S bond) in the sample containing one isolated Fe$_{Mo}$ point defect (fig. S11A). The decreased bond length enhances the crystal field in the trigonal prismatic coordination of S ligands, leading to the splitting of the *d*-orbitals of the Fe atom. As shown in the projected density of states (PDOS) in Fig. 4B, the flat band just below the $E_F$ is dominated by the Fe-$d_{z^2}$ orbital, while the degenerate Fe-$d_{xy}$ and Fe-$d_{x^2-y^2}$ orbitals mainly contribute to the flat band just above $E_F$. Two additional bands at ~0.75 eV above $E_F$ originate from the Fe-$d_{xz}$ and Fe-$d_{yz}$ orbitals. The presence of unoccupied bands above the $E_F$ suggests n-type doping in the monolayer MoS$_2$. By integrating PDOS around 0.75 eV, it is found that per Fe$_{Mo}$ point defect provides 1.49 electrons for monolayer MoS$_2$.

The band structure of monolayer MoS$_2$ with a 3Fe$_{Mo}$-S associate is shown in Fig. 4C. Increased Fe substitution introduces additional in-gap bands, and the $E_F$ intersecting these impurity bands indicates high electrical conductivity, consistent with the electrical transport results in Fig. 3D. Notably, the PDOS in Fig. 4D exhibits a sharp peak of ~195.0 states/eV near -0.58 eV above the VBM. This localized state can accommodate more electrons from valence band of monolayer MoS$_2$, which plays a crucial role to the hole doping. Theoretical calculations show that every 3Fe$_{Mo}$-S associate donates 11.14 holes to the system, exceeding the electron contribution from a single $Fe_{Mo}$ point defect by approximately an order of magnitude. This also explains why even in the monolayer $MoS_2$ system where the concentration of $3Fe_{Mo}$-S associates is much less than that of isolated $Fe_{Mo}$ point defects, the sample still features a degenerate p-type semiconductor.

**$3Fe_W$-S associates in monolayer Fe-doped $WS_2$**

In order to explore whether similar associates can exist stably in 2D semiconductors analogous to $MoS_2$ and how they affect the properties of these materials, monolayer Fe-doped $WS_2$ was synthesized (see Methods). The EDX spectrum of Fe-doped $WS_2$ flakes confirms the existence of Fe atoms in $WS_2$ (fig. S16). Figure 5A shows the atomic-resolution HAADF-STEM image of a monolayer Fe-doped $WS_2$. It can be seen that isolated $Fe_W$ point defect (white circle) and $3Fe_W$-S associate (yellow triangle) indeed existed in the sample. The enlarged HAADF-STEM images of one $Fe_W$ point defect marked with blue box and one $3Fe_W$-S associate marked with red box in Fig. 5A are displayed in Fig. 5 (B and C), respectively. The contrast intensity profile in Fig. 5D measured along the blue broken line in Fig. 5A confirms that the Fe atom substitutes the W site, and the contrast intensity profile in Fig. 5E measured along the red broken line in Fig. 5A confirms that there is only single S atom at the center of the $3Fe_W$-S associate. Similar to Fe-doped $MoS_2$, Fe substitutes W randomly and the Fe doping concentration in monolayer Fe-doped $WS_2$ is ~4.6 at. % (fig. S17). Monolayer CVD-grown $WS_2$ is a well n-type semiconductor as reported widely (*38, 39*). Figure 5 (F and G) shows the output and transfer curves of monolayer Fe-doped $WS_2$, showing good contact and a degenerate p-type property. This anomalous degenerate p-type semiconductor conductivity behavior suggests that $3Fe_W$-S associates contributing holes to monolayer $WS_2$, which is similar to that in monolayer Fe-doped $MoS_2$.

**Discussion**

In this work, $Fe_{Mo}$ point defects and $3Fe_{Mo}$-S associates are fabricated and identified in CVD grown-monolayer Fe-doped $MoS_2$. Combined STM/STS, electrical transport measurements, and first-principles calculations confirm that $Fe_{Mo}$ point defects act as electron donors, while $3Fe_{Mo}$-S associates serve as hole donors, demonstrating that the same element dopant can induce distinct electronic behaviors dependent on the structural configurations. Moreover, similar $3Fe_W$-S associates with hole doping properties are identified in monolayer Fe-doped $WS_2$, indicating the generality of this mechanism. Our research explains why Fe-doped $MoS_2$ samples can present different electrical properties in previous studies, and highlights the potential of defect associates for tailoring the characteristics of 2D semiconductors.

## Materials and Methods

### Sample fabrication

**Synthesis of monolayer Fe-doped $MoS_2$**: Monolayer Fe-doped $MoS_2$ flakes were fabricated using CVD method (fig. S1). A quartz boat containing 50 mg S powder (99.5%, Alfa Aesar A Johnson Matthey) was placed in the low temperature zone of the tubular furnace (upstream), and another quartz boat containing 1.5 mg NaCl (99.0%, Alfa Aesar A Johnson Matthey)/precursor mixture with a mass ratio of 9:1 was placed in the high temperature zone (downstream). Two kinds of precursors are used to prepare Fe-doped $MoS_2$, one is $Fe_2Mo_3O_{12}$ (98.0%, Alab) and the other is $MoO_3$ (99.95%, Aladdin)/$FeS_2$ (99.9%, Aladdin) mixture. A cleaned $Si/SiO_2$ (oxide layer thickness:285 nm) wafer was positioned atop the quartz boat within the high-temperature zone to serve as the growth substrate. Before heating the tube furnace, $N_2$ with a flow rate of 400 stand cubic centimeter per minute (sccm) was introduced into the tube furnace for 10 min along the direction from the low to the high temperature zone to exhaust the air, and then the $N_2$ flow rate was adjusted to 150 sccm. The thermal protocol comprised three stages: (i) Ramping the high-temperature zone to 600 °C at 20 °C/min while initiating S powder heating in the low-temperature zone; (ii) Maintaining thermal gradients until the low-temperature zone reached 160 °C (achieved within 5 min) coinciding with high-temperature zone stabilizing at 700 °C, followed by 10 min isothermal growth for Fe-doped $MoS_2$ synthesis; (iii) Naturally cooling the furnace to ambient temperature. The Fe doping concentration in monolayer $MoS_2$ can be tuned by controlling the mole ratio between $FeS_2$ and $MoO_3$ in precursor.

**Synthesis of monolayer pristine $MoS_2$**: It is the same with the synthesis of monolayer Fe-doped $MoS_2$ except that the precursor is replaced by pure $MoO_3$.

**Synthesis of monolayer Fe-doped $WS_2$**: It is the same with the synthesis of monolayer Fe-doped $MoS_2$ except that the precursor is replaced $FeSO_4 \cdot 7H_2O$ (≥99.0%, Xilong) /$WO_3$ (99.9%, Aladdin) mixtrue with 60 at.% Fe, and the growth is carried out at 800 °C for 3 min.

### Atomic structure characterization of samples

All samples are transferred from the growth to the test substrates via water-assisted transfer method (*40*). All the TEM experiments were completed on the condenser Cs-STEM (FEI, Titan$^3$ Themis Z) with 300 kV accelerating voltage.

### STM/STS measurements

The monolayer Fe-doped $MoS_2$ flakes stacking on Au substrate were transferred into an ultrahigh vacuum (UHV) chamber of the STM system (USM-1300 from UNISOKU) and degassed at 300 °C for 4-6 hours with a pressure better than $3\times10^{-10}$ Torr. The degassed sample was then transferred into a STM scanning chamber for *in situ* measurement in the UHV and low-temperature environment. All the STM measurements were performed at the constant-current mode and 77 K using an electrochemically etched tungsten tip. The d$I$/d$V$ spectra were measured using a lock-in technique with a bias modulation amplitude of $V_{mod}$ = 10 mV and frequency of 973.1 Hz.

### Device fabrication

The FET devices were fabricated by electron beam lithography, electrodes deposition and lift-off processes. The electrodes are Ta (10 nm)/Pt (40 nm) deposited by magnetron sputtering.

**First-principles calculations**

The electronic structures of monolayer pristine and Fe-doped MoS$_2$ and the total energy calculations of all Fe-related configurations were performed by Vienna ab initio simulation package (VASP) (*41, 42*). The cutoff energy of 520 eV is taken for the plane-wave basis set (*43*). The exchange-correlation interactions are described by the generalized gradient approximation (GGA) with Perdew-Burke-Ernzerhof (PBE) type functional (*44*). Atoms are fully relaxed until the Hellmann-Feynman forces on each atom are less than 0.01 eV/Å. The first Brillouin zone sampling was performed using Γ-centered Monkhorst-Pack grids: a 3×3×1 mesh for lattice relaxation of the 4×4×1 supercell, followed by a higher-density 5×5×1 mesh for electronic structure determination. To mimic a 2D material, a layer distance of 20 Å along *z* direction is used to eliminate the interference between periodic images. The energy convergence threshold is established at $10^{-6}$ eV. Post-processing of data is carried out using VASPKIT code (*45*). Based on the force matrices from the first-principles calculations, the phonon modes of monolayer MoS$_2$ are carried out by the PHONOPY code (*46, 47*), where all atomic positions were fully optimized until the residual forces were less than 0.0001eV/Å.

**Supplementary Materials**
**This PDF file includes:**
Supplementary Text
Figs. S1 to S18
Table S1
References

**Acknowledgements**
We thank Xuegang Chen and Mingsheng Long at Anhui University for their assistance in electrical characterizations.

**Funding:**
This work was supported by the Natural Science Foundation of Gansu Province (24JRRA398), the Excellent Doctoral Student Program of Gansu Province (23JRRA1130 and 25JRRA741), the Open Fund of Information Materials and Intelligent Sensing Laboratory of Anhui Province (IMIS202301), the Project of Yunnan Key Laboratory of Electromagnetic Materials and Devices, Yunnan University (ZZ2024007), the Gansu Youth Science and Technology Fund (23JRRA624), and the National Natural Science Foundation of China (62104089, 12274185, 12474128,12204008 and 12104004).

**Author contributions:** L.Z. and Y.P. conceived the whole project. X.Q., J.Z. and L.Z. designed the experiments, prepared and characterized the samples, analyzed and interpreted the main data. X.Q., X.Y., Z.Z, and L.S. finished the STM/STS measurements and analyzed corresponding data. X.Q., X.P., H.M. D.G., A.Z. and M.S. performed the Raman spectra collection, DFT and first-principles calculations. C.Y., H.Z. and Y.W. assisted in the electrical characterizations of samples. H.L. assisted in the AFM characterizations. X.Q., L.Z., Z.Z., H.M., M.S. and Y.P. co-wrote the manuscript, and all authors commented on the manuscript.

**Competing interests:** The authors declare no competing interests.

**Data and materials availability:** The data that support the findings of this study are available from the corresponding authors upon reasonable request.

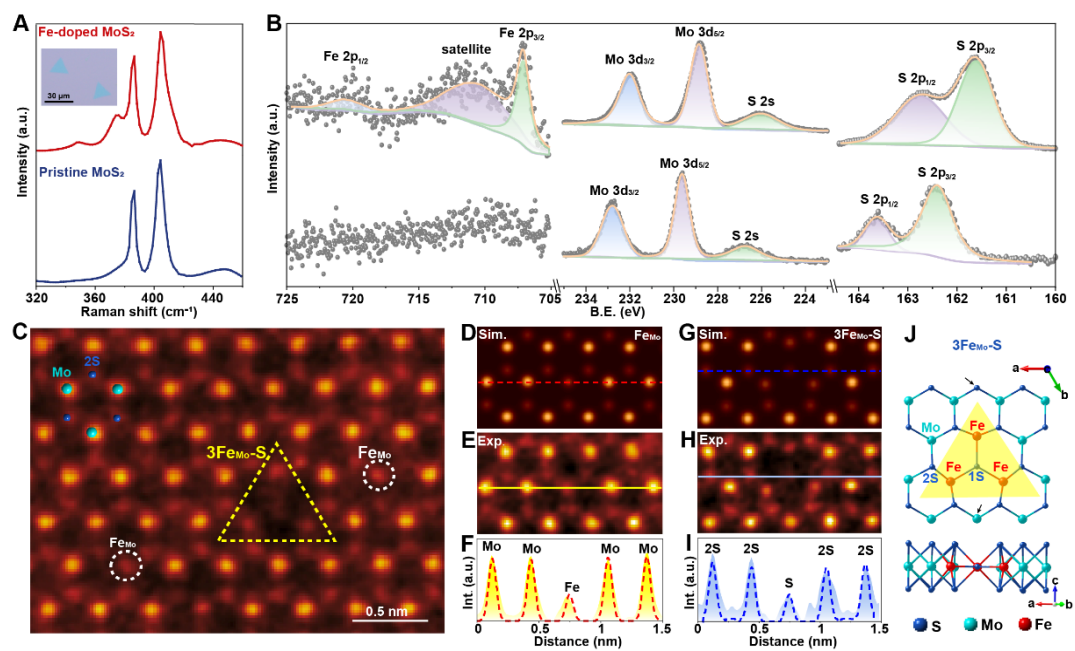

**Fig. 1. Characterizations of monolayer Fe-doped and pristine MoS$_2$.** (**A**) Raman spectra of monolayer Fe-doped and pristine MoS$_2$, the inset in (**A**) shows the optical image of freshly prepared monolayer Fe-doped MoS$_2$ flakes. (**B**) Fine scanning Fe 2p, Mo 3d, and S 2p XPS spectra of Fe-doped and pristine MoS$_2$. (**C**) Atomic scale HAADF-STEM image of monolayer Fe-doped MoS$_2$, white circle and yellow triangle respectively indicate Fe$_{Mo}$ point defect and 3Fe$_{Mo}$-S associate. The atomic images (**D** and **G**) simulated by QSTEM software and (**E** and **H**) taken experimentally. (**F** and **I**) The contrast intensity curves measured along the broken and solid lines marking the atomic positions in (**D** and **E**) and (**G** and **H**) respectively. (**J**) Schematic diagram of the atomic structure of one 3Fe$_{Mo}$-S defect associate, and the yellow triangle marks the main part of the associate. Blue, cyan, and red spheres denote S, Mo, and Fe atoms, respectively. The S and Mo atoms marked with arrows in the top view are hidden in the side view in order to observe the 3Fe$_{Mo}$-S configuration more clearly.

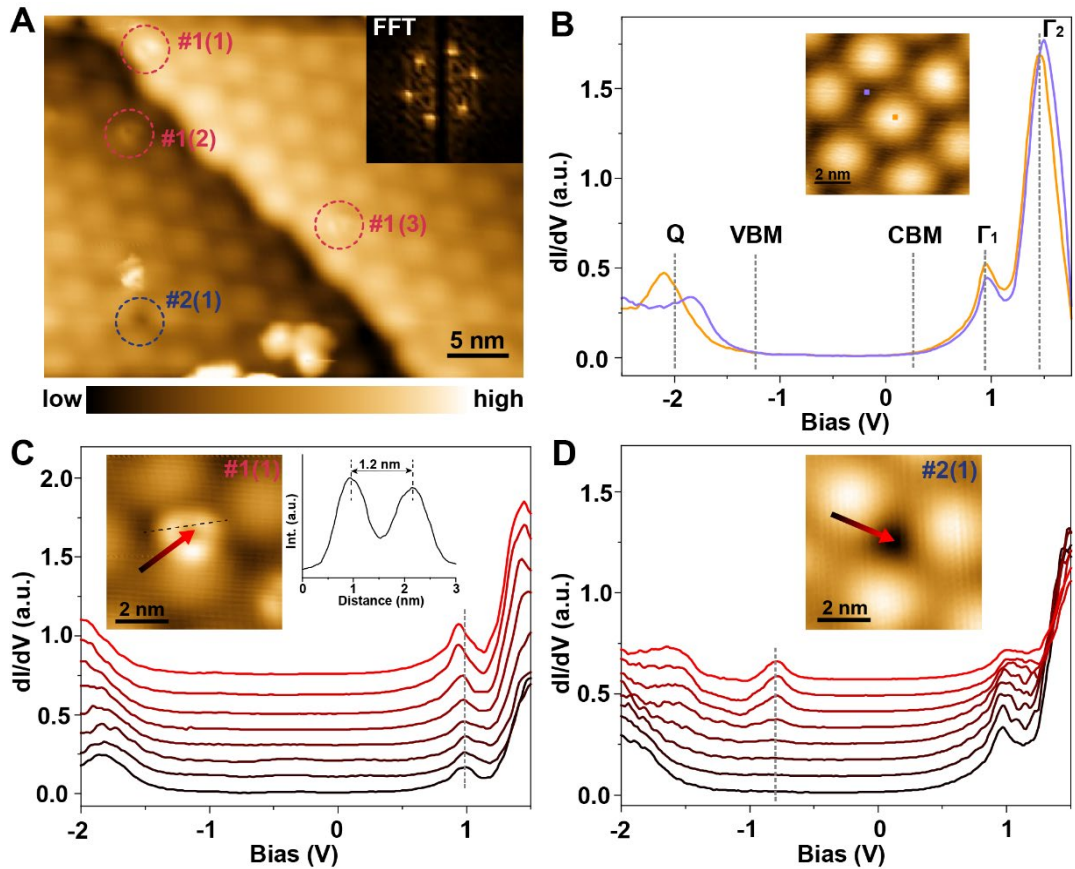

**Fig. 2. STM/STS characterizations of isolated Fe$_{Mo}$ point defects and 3Fe$_{Mo}$-S associates.** (**A**) A STM image of monolayer Fe-doped MoS$_2$ stacking on Au (111) substrate and its corresponding FFT image (inset) ($V_{set}$ = -0.8 V, $I_{set}$ = 30 pA). (**B**) dI/dV spectra measured on the top (orange point) and hollow (purple point) regions on pristine MoS$_2$ monolayer. A series of STS spectra measured along the arrow direction approaching the center of (**C**) an isolated Fe$_{Mo}$ point defect and (**D**) a 3Fe$_{Mo}$-S associate. Spectra are vertically shifted for clarity. The curve in the right inset of (**C**) is the contrast intensity curve measured along the black broken line in the left inset of (**C**). All spectra are measured at the setpoint of $V_{set}$ = -2 V, $I_{set}$ = 300 pA and at the temperature of $T$ = 77 K.

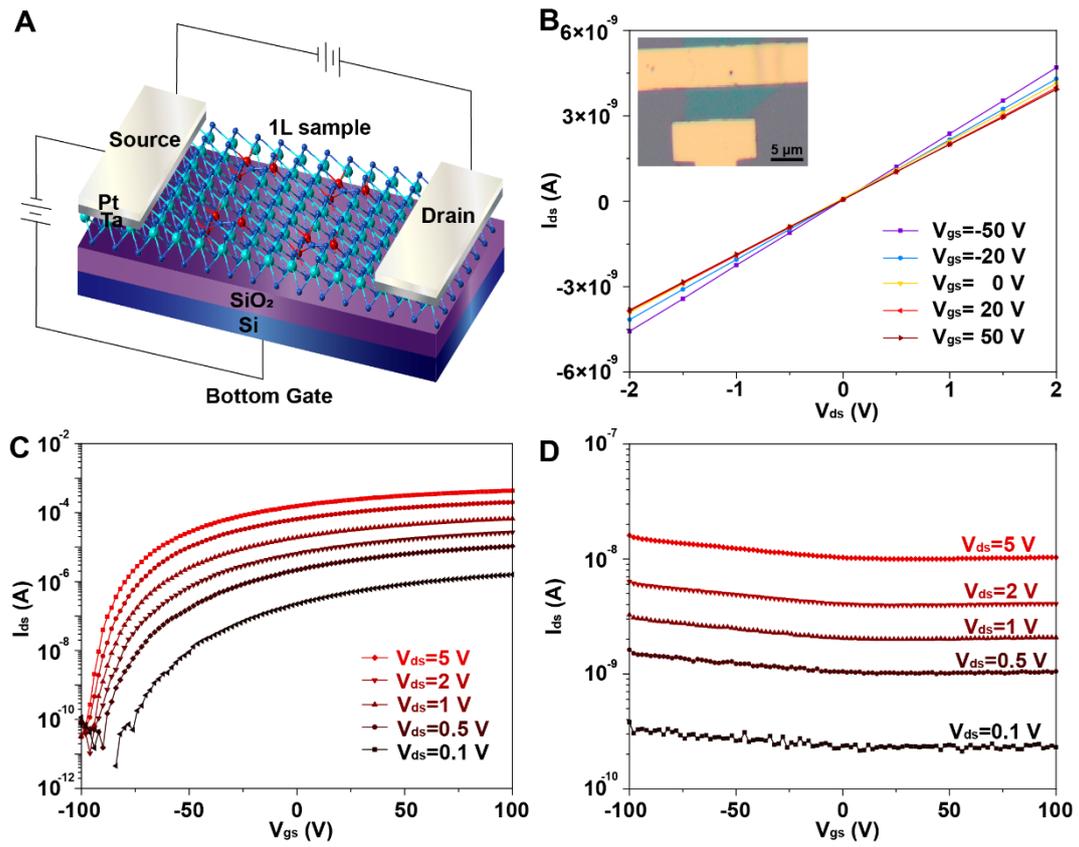

**Fig. 3. Electrical transport properties of monolayer pristine and Fe-doped MoS$_2$.** (**A**) Schematic diagram of the monolayer Fe-doped MoS$_2$ FET device. (**B**) The output curves of monolayer Fe-doped MoS$_2$ and the optical image of a practical device (inset). The transfer curves of monolayer (**C**) pristine MoS$_2$ and (**D**) Fe-doped MoS$_2$.

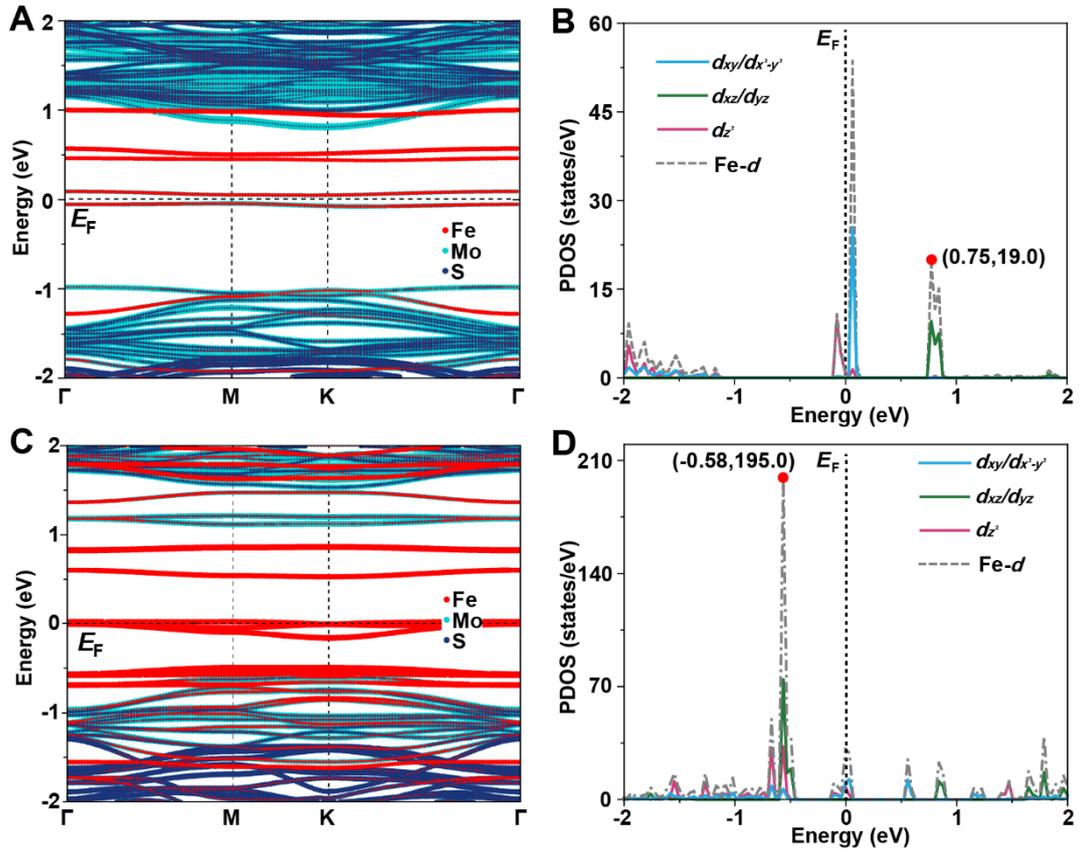

**Fig. 4. Calculated electronic structures of monolayer pristine and Fe-doped MoS$_2$.** Band structure of monolayer MoS$_2$ (**A**) containing one isolated Fe$_{Mo}$ point defect and (**C**) containing one 3Fe$_{Mo}$-S associate. The red, cyan, and dark blue dots represent the contribution of Fe, Mo, and S atoms in monolayer Fe-doped MoS$_2$, respectively. PDOS of monolayer MoS$_2$ (**B**) containing one isolated Fe$_{Mo}$ point defect and (**D**) containing one 3Fe$_{Mo}$-S associate. The blue, green, and pink solid lines represent the contribution of $d_{xy}/d_{x^2-y^2}$, $d_{xz}/d_{yz}$ and $d_{z^2}$ orbits of Fe in Fe-doped MoS$_2$, respectively, and the gray dotted line indicates their overall contribution.

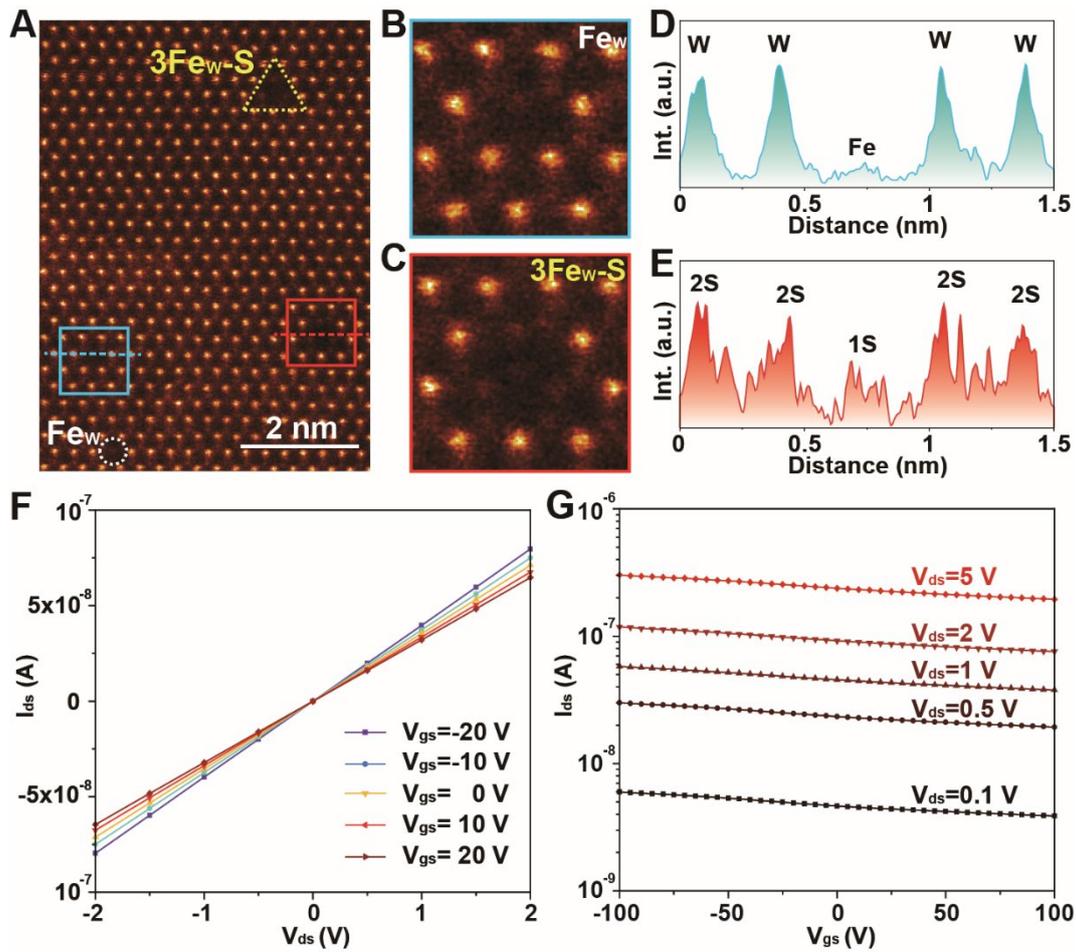

**Fig. 5. Atomic structure and electrical transport properties of monolayer Fe-doped WS$_2$.** (**A**) Atomic-scale HAADF-STEM image of monolayer Fe-doped WS$_2$. Two kinds of defects: (**B**) isolated Fe$_W$ piont defect and (**C**) 3Fe$_W$-S defect associate enlarged from the positions of blue and red rectangle in (**A**). (**D**), The intensity profile of the atomic positions marked with blue broken line of Fe and its nearby W site at the blue rectangle in (**A**). (**E**) The intensity profile of the atomic positions marked with red broken line of S and its nearby S$_2$ site at the red rectangle in (**A**). (**F**) The output curves and (**G**) the transfer curves of monolayer Fe-doped WS$_2$.

# Supplementary Materials for

**Origin of anomalous p-type conductivity in monolayer Fe-doped MoS$_2$**

Xiangning Quan *et al.*

*Lili Zhang, zll@lzu.edu.cn; Anmin Zhang, amzhang@lzu.edu.cn; Zongyuan Zhang, zongyuanzhang@ahu.edu.cn; Yong Peng, pengy@lzu.edu.cn

**This PDF file includes:**

Supplementary Text
Figs. S1 to S17
Table S1
References 48 and 49

**Supplementary Text**
**1. Collection and analysis of the Raman spectra of monolayer Fe-doped MoS$_2$**

Raman spectra of monolayer Fe-doped MoS$_2$ were collected with a LABRAM HR Evolution system at room temperature, which is equipped with a single grating of 800 mm focus length and liquid-nitrogen-cooled CCD. A laser with a wavelength of 532 nm was focused on the surface of monolayer pristine and Fe-doped MoS$_2$ flakes, and the excitation laser power was kept less than 0.3 mW to avoid heating effect. Here, we use the Porto notation $k_i$ ($e_i$, $e_s$) $k_s$, where $k_i$ and $k_s$ denote the direction of incident and scattered light and ($e_i$, $e_s$) are their polarization, respectively. x, y and z correspond to the laboratory Cartesian coordinate system, with the z axis parallel to the c axis of sample. An angle-dependent Raman was performed with the polarization configuration $z(yy)\bar{z}$ and the z direction as the axis. The sample was rotated from 0° to 180° by taking 10° as the step to record the Raman peak intensities, complete all results from 0° to 360° by angle symmetry.

In Raman spectroscopy, the Raman scattering intensity can be expressed as $I \propto |\vec{e_i} \cdot \vec{M} \cdot \vec{R} \cdot \vec{M}^{-1} \cdot \vec{e_s}|^2$, where $\vec{e_i}$ is the polarization vector of the incident light, $\vec{e_s}$ is the polarization vector of the scattered light, and $\vec{M}$ is the transformation matrix relating with the crystal coordinate to the laboratory coordinate. If the angle between the two coordinate systems is $\theta$, $\vec{M}$ can be espressed as $\vec{M} = \begin{pmatrix} cos\theta & sin\theta & 0 \\ -sin\theta & cos\theta & 0 \\ 0 & 0 & 1 \end{pmatrix}$ in our experiment. The Raman tensor $\vec{R}$ is determined by the symmetry of the crystal. Therefore, by rotating the sample and performing Raman measurements at regular angular intervals, the function relationship between the integrated Raman intensity and $\theta$ can be used to determine the Raman tensor and identify the Raman peaks. Figure S3A shows the polarized Raman spectra of monolayer Fe-doped MoS$_2$. Besides the two main first-order bands E' mode (in-plane vibration of S and Mo atoms, referred as $E^1_{2g}$ in even layer sample or bulk) and A$_1$' mode (out-of-plane vibration of S atoms, referred as A$_{1g}$ in even layer sample or bulk) around 385.1 cm$^{-1}$ and 404.5 cm$^{-1}$, Raman peaks also arise at 349.4 cm$^{-1}$, 375.8 cm$^{-1}$, and around 455 cm$^{-1}$. The Raman peak locating at

455 cm$^{-1}$ is commonly called the 2LA band of MoS$_2$ (*48*). Monolayer pristine MoS$_2$ has six optical phonon modes, the frequencies of the other four optical phonon modes are significantly away from 349.4 cm$^{-1}$ and 375.8 cm$^{-1}$, which are marked as P$_1$ and P$_2$ later, respectively. To identify P$_1$ and P$_2$, phonon spectra calculations and symmetry analysis were performed on monolayer pristine MoS$_2$ as shown in fig. S3B and Table S1.

Angle-resolved polarized Raman spectra were performed on monolayer Fe-doped MoS$_2$ samples under $z(yy)\bar{z}$ polarization configuration and the results are shown in fig. S3C. The integrated Raman intensities of P$_1$, P$_2$, E' and A$_1$' modes under the $z(yy)\bar{z}$ polarization configuration were fitted with Lorentzian curve, and the results are shown in fig. S3 (D to G). For E' and A$_1$' displayed in fig. S3 (D and E), the integrated intensities are approximately isotropic at all polarization angles, consistent with their Raman tensors corresponding to their irreducible representations. The integrated intensities of P$_1$ and P$_2$ are also approximately isotropic as shown in fig. S3 (F and G), which corresponds to the Raman tensors of A$_1$'(K) and A$_1$(M), respectively (when the *xx* and *yy* diagonal elements of the A$_1$(M) Raman matrix are approximately equal). Combined with the calculated phonon frequencies at the M and K points, P$_1$ can be attributed to A$_1$(M) at 340.1 cm$^{-1}$, and P$_2$ can be assign to A$_1$(M) or A$_1$'(K) at 390.7 cm$^{-1}$.

**2. The total energy of 3Fe$_{Mo}$-S associate with the position of central S atom.**

It is found that the position of S atom along the normal line of the equilateral triangle formed by three Fe atoms plays a key role in determining the total energy of 3Fe$_{Mo}$-S associate. Setting the total energy of the system to be 0 eV upon S at central position, the total energy difference (ΔE) shown in fig. S10 demonstrates that ΔE is changed with the S position along the normal line of the triangle. ΔE gets the minimum when the central S atom is coplanar with its neighboring Fe atoms, suggesting that 3Fe$_{Mo}$-S associate prefers to be a planar structure.

**3. The calculation method for Fe doping concentration.**

Figure S7 (A and D) shows the atomic HADDF-STEM image of monolayer pristine and Fe-doped $MoS_2$ along the [0001] direction, respectively. All enlarged atomic diagrams in Fig. 1 (C, E and H) in the main text are taken from the corresponding positions in fig. S7D and are marked with white, yellow, and blue boxes, respectively. All Mo sites and Fe sites in fig. S7 (A and D) are labelled as cyan and red circles respectively by CalAtom software (*49*) according to the Z-contrast of different atoms, and the results are shown in fig. S7 (B and E), respectively. The $Fe_{Mo}$ point defects and the $3Fe_{Mo}$-S associates can be clearly seen in fig. S7E and the $3Fe_{Mo}$-S associates are marked with red triangles. After analyzing the contrast intensity distribution of all labelled spots in fig. S7E, the contrast intensity distribution splits into two Gaussian peaks. One is located at lower intensity range and the other locates at higher intensity range, which correspond to Fe sites and Mo sites, respectively. By calculating the number of atoms belonging to these two peaks with different brightness, the doping concentration of Fe can be determined to be 6.9 at.%. Accordingly, the analyzing result of fig. S7B only shows one peak as shown in fig. S7C since there are only Mo atoms in the metal sites.

# 4. Calculation of the formation energy for different Fe-related defect configurations in monolayer Fe-doped MoS$_2$.

Four different Fe-related defect configurations are considered, which are named Fe$_{Mo}$ point defect, 3Fe$_{Mo}$-S associate, 3Fe$_{Mo}$ (I) and 3Fe$_{Mo}$ (II), respectively. 3Fe$_{Mo}$ (I) represents the configuration of 3Fe$_{Mo}$ associate with two S atoms still remain in the center, and 3Fe$_{Mo}$ (II) represents the configuration of 3Fe$_{Mo}$ associate consisting of three adjacent Fe atoms without central S atoms. Figure S11 shows the top views and side views of these four defect configurations.

The formation energies of above defect configurations are calculated using the following formula:

$$E_f = E_{def} - E_{host} + \sum_i n_i u_i,$$

where $E_{def}$ is the total energy of Fe-doped MoS$_2$ and $E_{host}$ is that of host MoS$_2$. Both energies are calculated based on 4×4×1 supercell. $n_i$ is the number of increase ($n_i < 0$) or decrease ($n_i > 0$) atom, and $u_i$ denotes the chemical potential of atomic number $n_i$. Naturally, we can respectively write out the concrete expressions of the formation energies for these four defect configurations:

Fe$_{Mo}$:
$$E_f = E_{def} - E_{host} + u_{Mo} - u_{Fe}$$

3Fe$_{Mo}$-S:
$$E_f = E_{def} - E_{host} + 3u_{Mo} + u_S - 3u_{Fe}$$

3Fe$_{Mo}$ (I):
$$E_f = E_{def} - E_{host} + 3u_{Mo} - 3u_{Fe}$$

3Fe$_{Mo}$ (II):
$$E_f = E_{def} - E_{host} + 3u_{Mo} - 3u_{Fe}.$$

Under S-rich condition, $u_{Fe}$ is half of the energy calculated in Fe cubic bulk, i.e. $u_{Fe} = 0.5E_{Fe}$, and $u_s = 0.125E_{S_8}$ with $E_{S_8}$ being the energy of one S$_8$ molecule. Then, $u_{Mo} = E_{MoS_2} - 2u_s$, where $E_{MoS_2}$ is the total energy of MoS$_2$ conventional unit cell. Thus, the formation energies of Fe$_{Mo}$ point defect, 3Fe$_{Mo}$-S associate, 3Fe$_{Mo}$ (I) and 3Fe$_{Mo}$ (II) are 0.10 eV, 1.03 eV, 2.52 eV and 3.75 eV, respectively, which are shown in fig. S12.

**Fig. S1.**

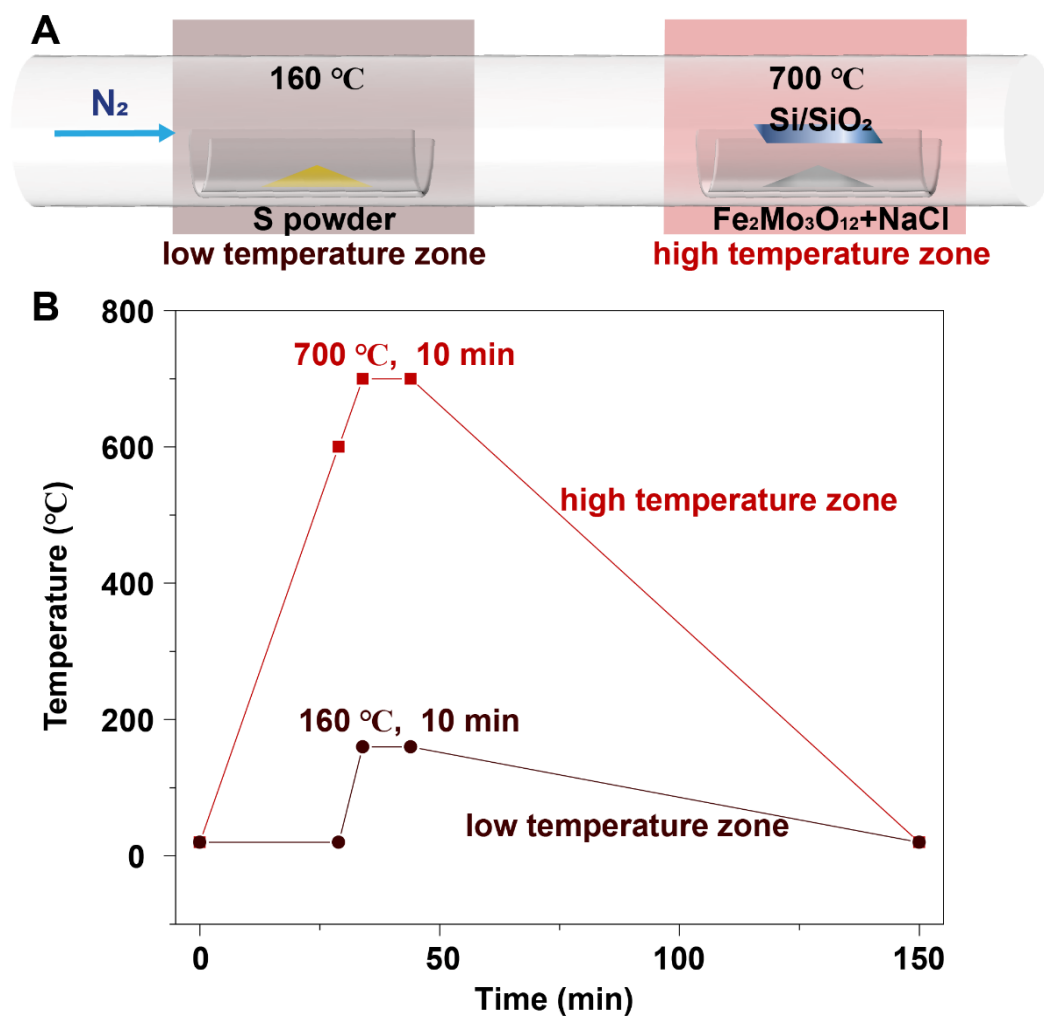

**Fig. S1. Synthesis of monolayer Fe-doped MoS$_2$.** (**A**) Schematic diagram for CVD fabrication of monolayer Fe-doped MoS$_2$. (**B**) Time-temperature control curve of each temperature zone in the synthesis process.

**Fig. S2.**

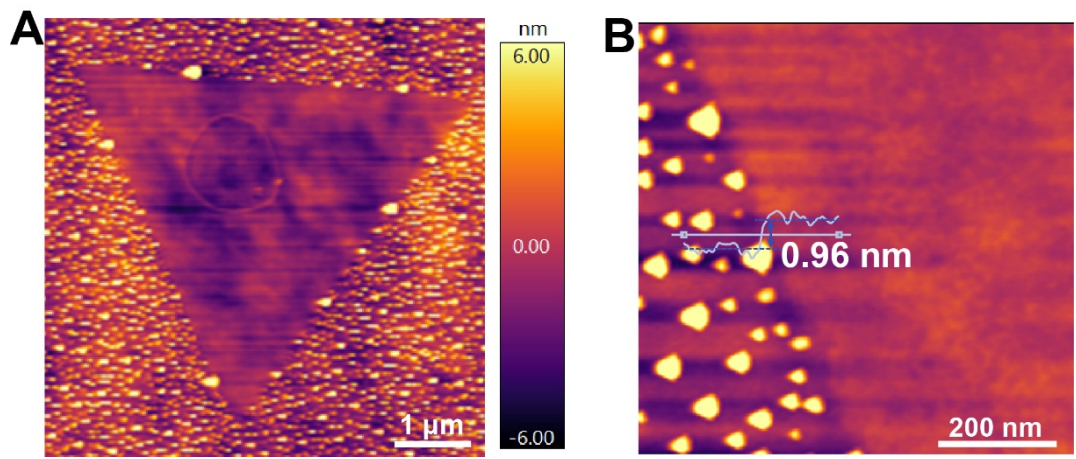

**Fig. S2. Atomic force microscopy (AFM) characterization of monolayer Fe-doped MoS$_2$.** (**A**) AFM image of a freshly prepared monolayer Fe-doped MoS$_2$ flake. (**B**) The height profile along the white line across the flake edge measured by AFM.

Fig. S3.

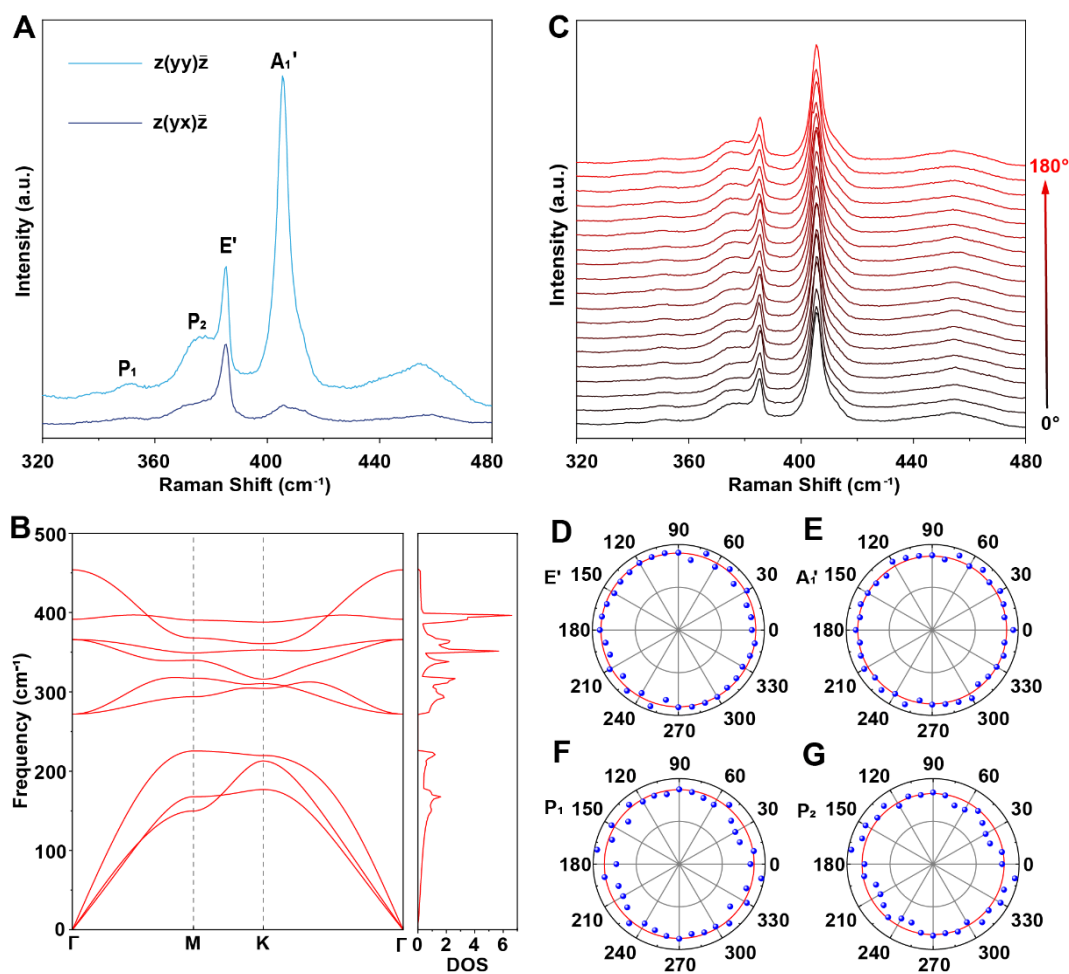

**Fig. S3. Raman spectra of monolayer Fe-doped MoS$_2$.** (**A**) Polarized Raman spectra of monolayer Fe-doped MoS$_2$. (**B**) Phonon dispersion and vibrational density of states (DOS) of monolayer MoS$_2$ calculated from density functional theory (DFT) calculation. (**C**) Angle-resolved polarized Raman spectra of monolayer Fe-doped MoS$_2$ from 0° to 180°. (**D** to **G**) Polar plots for the spectrally integrated intensities of E', A$_1$', P$_1$, and P$_2$ phonons as a function of incident polarization angle $\theta$.

**Fig. S4.**

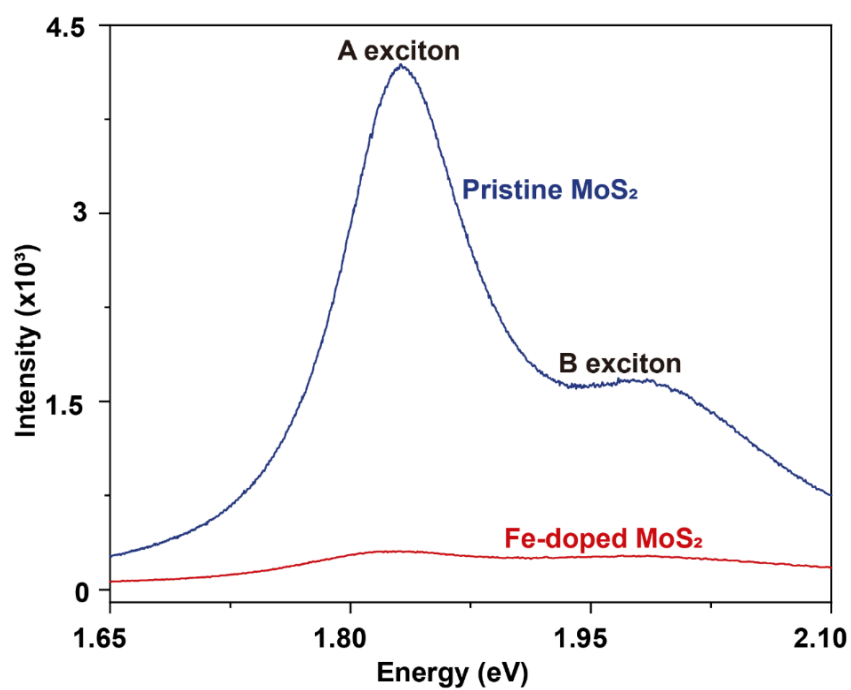

**Fig. S4. Photoluminescence (PL) spectra of monolayer Fe-doped and pristine MoS₂.** Pristine MoS$_2$ displays two peaks at 1.83 eV (A exciton) and 1.98 eV (B exciton). Incorporation of Fe atoms leads to a strong quenching effect of PL in monolayer Fe-doped MoS$_2$.

**Fig. S5.**

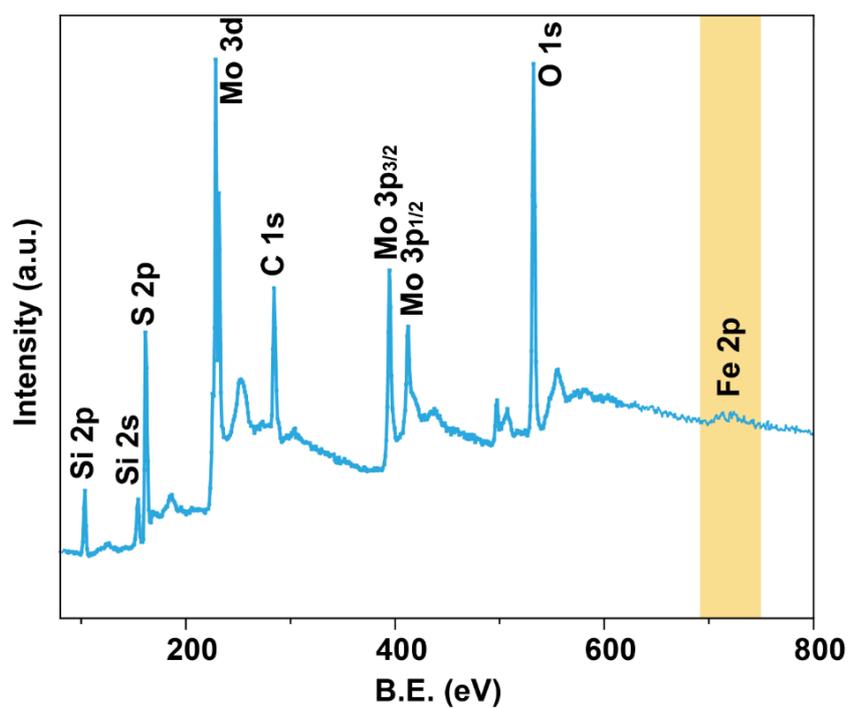

**Fig. S5. Survey XPS spectrum of as-grown Fe-doped MoS$_2$ on Si/SiO$_2$ substrate.** The XPS peaks of Si (103.4 eV, 153.3 eV) and O (532.9 eV) come from the Si/SiO$_2$ substrate.

**Fig. S6.**

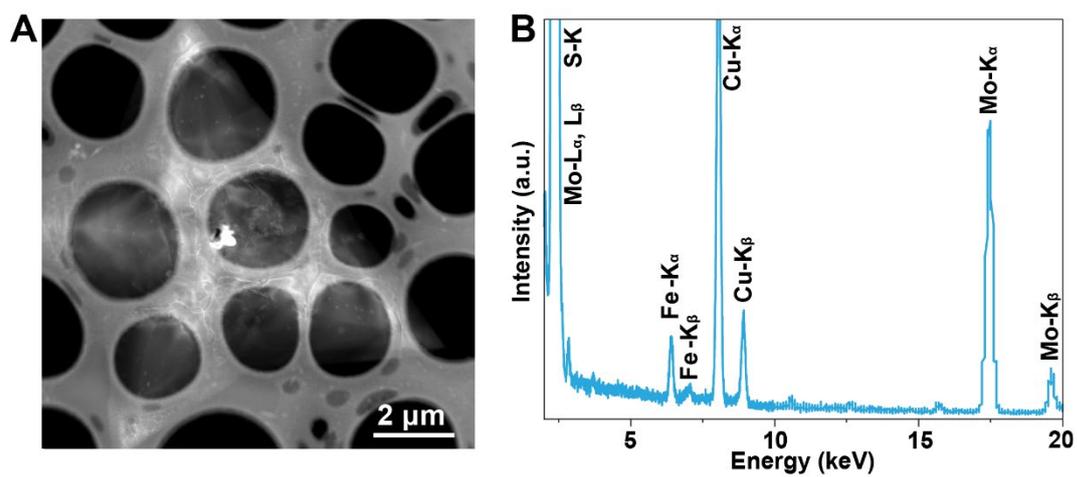

**Fig. S6.** (**A**) The HAADF-STEM image of a Fe-doped $MoS_2$ flake. (**B**) EDX spectrum of Fe-doped $MoS_2$ measured from sample in (**A**). The Cu signal comes from the TEM Cu grid.

Fig. S7.

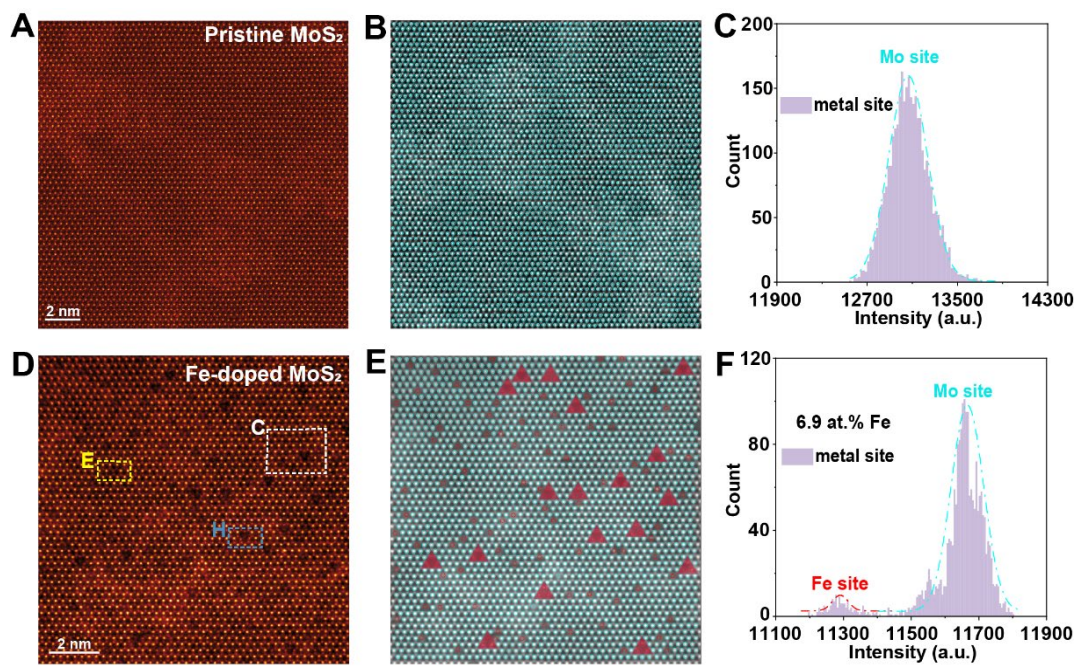

**Fig. S7. Microscopic characterizations of monolayer pristine and Fe-doped MoS₂.** Atomic-scale HAADF-STEM images of monolayer (**A**) pristine and (**D**) Fe-doped MoS₂. The labelled sites of Mo atoms (cyan circles) and Fe atoms (red circles) for monolayer (**B**) pristine and (**E**) Fe-doped MoS₂. (**C** and **F**) The statistical charts of atomic intensity for labelled atoms in monolayer (**B**) pristine and (**E**) Fe-doped MoS₂. The white, yellow, and blue boxes in (**D**) are the interception positions of the enlarged atomic structure presented in Fig. 1 (C, E and H) in the main text, respectively. 3Fe$_{Mo}$-S associates in (**E**) are marked with red triangles, and it can be seen that Fe atoms substitute Mo atoms randomly.

**Fig. S8.**

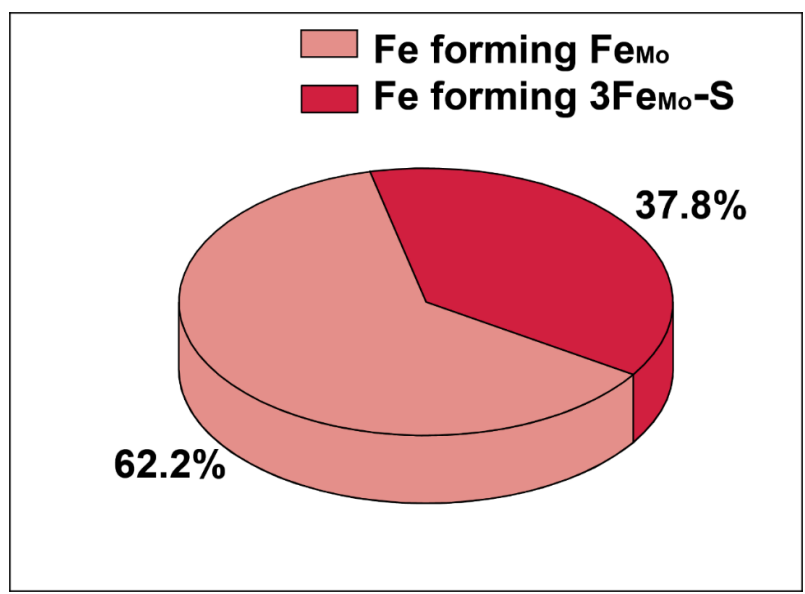

**Fig. S8. The proportions of Fe atoms in different defect types in monolayer Fe-doped MoS$_2$ synthesized from Fe$_2$Mo$_3$O$_{12}$ precursor.**

**Fig. S9.**

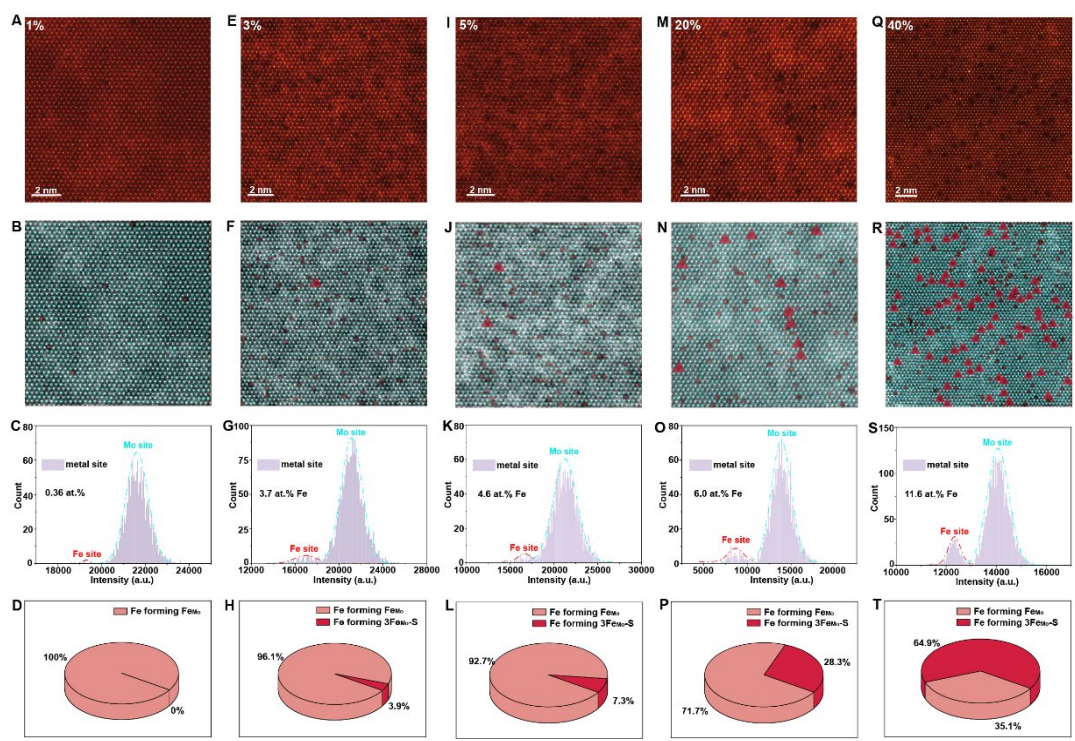

**Fig. S9. Statistical analysis of Fe doping concentration in monolayer Fe-doped MoS$_2$ synthesized from FeS$_2$/MoO$_3$ mixture precursor with different mole ratios of Fe.** Atomic HAADF-STEM images of monolayer Fe-doped MoS$_2$ grown with FeS$_2$ accounting for (**A**) 1 at. %, (**E**) 3 at. %, (**I**) 5 at. %, (**M**) 20 at. %, and (**Q**) 40 at. % in FeS$_2$-MoO$_3$ precursor. Metallic atom mapping of monolayer Fe-doped MoS$_2$ grown with FeS$_2$ accounting for (**B**) 1 at. %, (**F**) 3 at. %, (**J**) 5 at. %, (**N**) 20 at. %, and (**R**) 40 at. % in FeS$_2$-MoO$_3$ precursor, in which the positions of 3Fe$_{Mo}$-S associates are marked with red triangles. The cyan and red cycles are Mo sites and Fe sites, respectively. Statistical charts of atomic intensity for labelled atoms of monolayer Fe-doped MoS$_2$ grown with FeS$_2$ accounting for (**C**) 1 at. %, (**G**) 3 at. %, (**K**) 5 at. %, (**O**) 20 at. %, (**S**) 40 at. % in FeS$_2$-MoO$_3$ precursor and the proportions of Fe atoms in different defect types in monolayer Fe-doped MoS$_2$ grown with FeS$_2$ accounting for (**D**) 1 at. %, (**H**) 3 at. %, (**L**) 5 at. %, (**P**) 20 at. %, (**T**) 40 at. % in FeS$_2$-MoO$_3$ precursor. The red and pink areas in the pie chart represents the contents of 3Fe$_{Mo}$-S associate and isolated Fe$_{Mo}$ point defect, respectively.

**Fig. S10.**

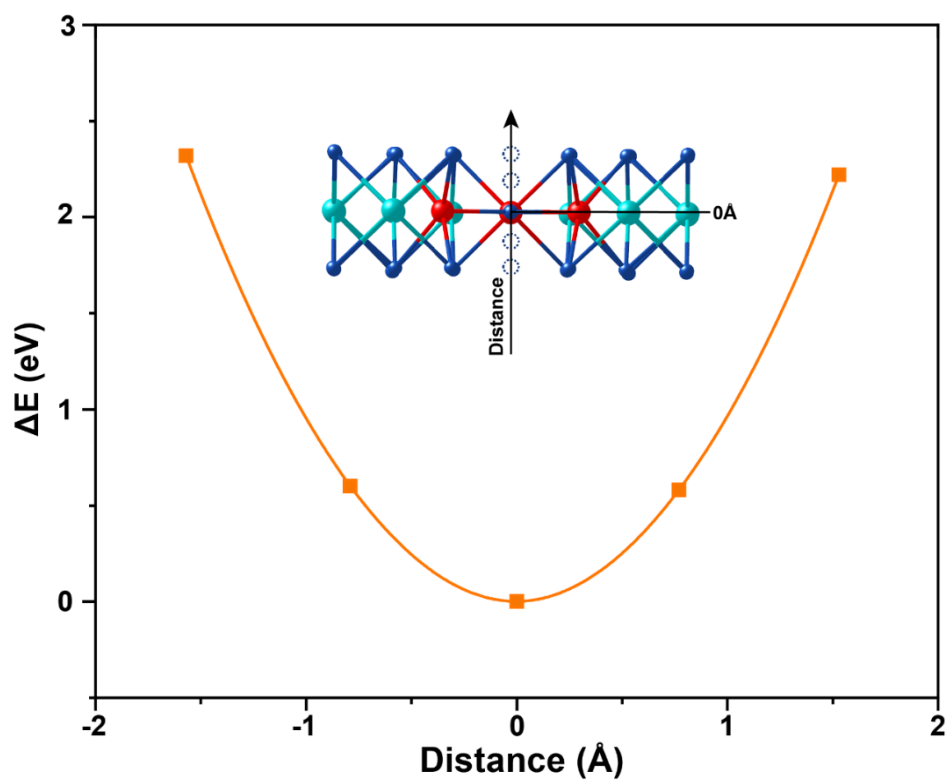

**Fig. S10. The total energy difference (ΔE) of 3Fe$_{Mo}$-S associate with the position of the central S atom.** Assuming that the plane of Fe atoms is the zero point of the coordinate axis, the moving distance of S atom is -1.57, -0.79, 0, 0.77 and 1.53 Å from the bottom to top. Set the ΔE of the system as 0 eV when the position of S atom is in the plane of Fe atoms, and the ΔE of the system from the bottom to top is 2.32, 0.6, 0, 0.58 and 2.22 eV, respectively.

Fig. S11.

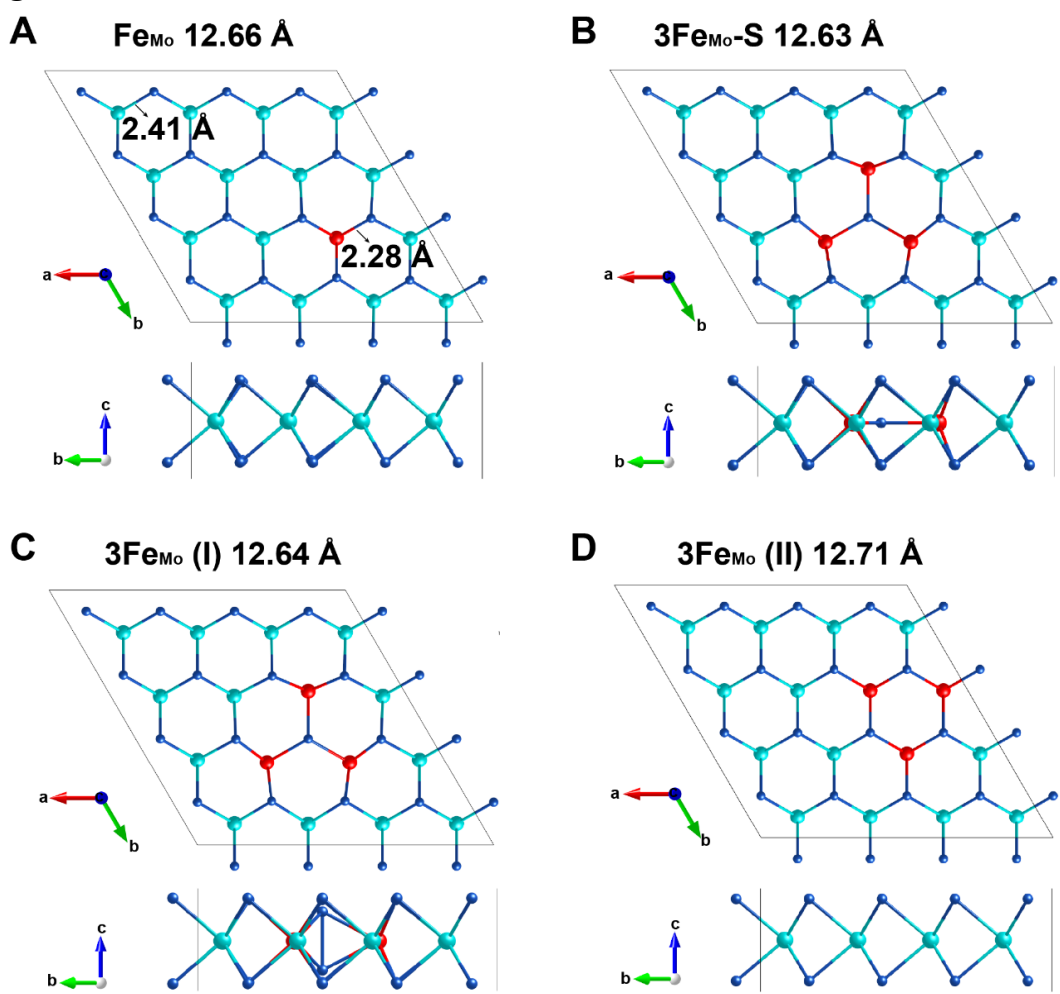

**Fig. S11. The unit structures of 4×4×1 supercell for monolayer Fe-doped MoS$_2$ with different Fe-containing configurations:** (**A**) Fe$_{Mo}$, (**B**) 3Fe$_{Mo}$-S, (**C**) 3Fe$_{Mo}$ (I), (**D**) 3Fe$_{Mo}$ (II). Plots display top (top panel) and side (bottom panel) views of theses configurations. The corresponding chemical bond length is indicated by arrows in (**A**). The lattice parameters of every configuration after relaxation are also outlined. Blue, cyan, and red spheres denote S, Mo, and Fe atoms, respectively. The length marked in every figure is the cell parameter (side length of diamond border) of each 4×4×1 supercell.

**Fig. S12.**

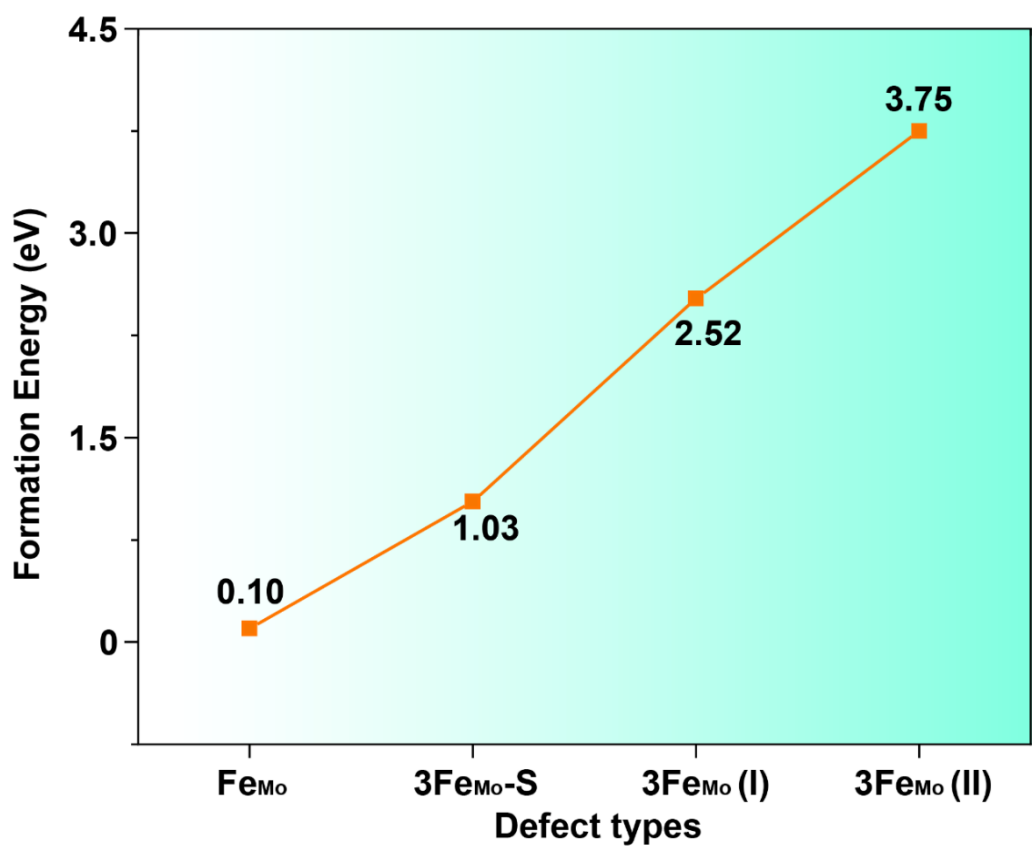

**Fig. S12.** The formation energy of Fe$_{Mo}$ point defect, 3Fe$_{Mo}$-S associate, 3Fe$_{Mo}$ (I) and 3Fe$_{Mo}$ (II) under S-rich condition.

**Fig. S13.**

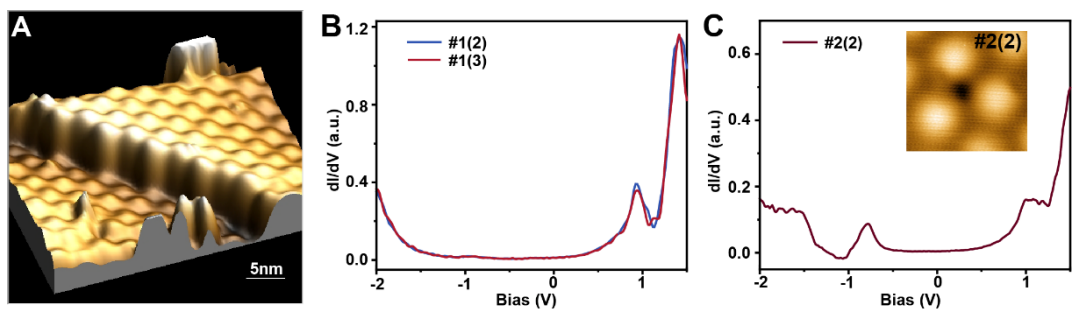

**Fig. S13.** (**A**) Corresponding 3D image of the atomic structure shown in Fig. 2A in the main text. dI/dV spectra in the center of (**B**) defect #1(2) and defect #1(3) as marked in Fig. 2A in the main text. (**C**) dI/dV spectrum in the center of defect #2(2) located at another position (inset).

**Fig. S14.**

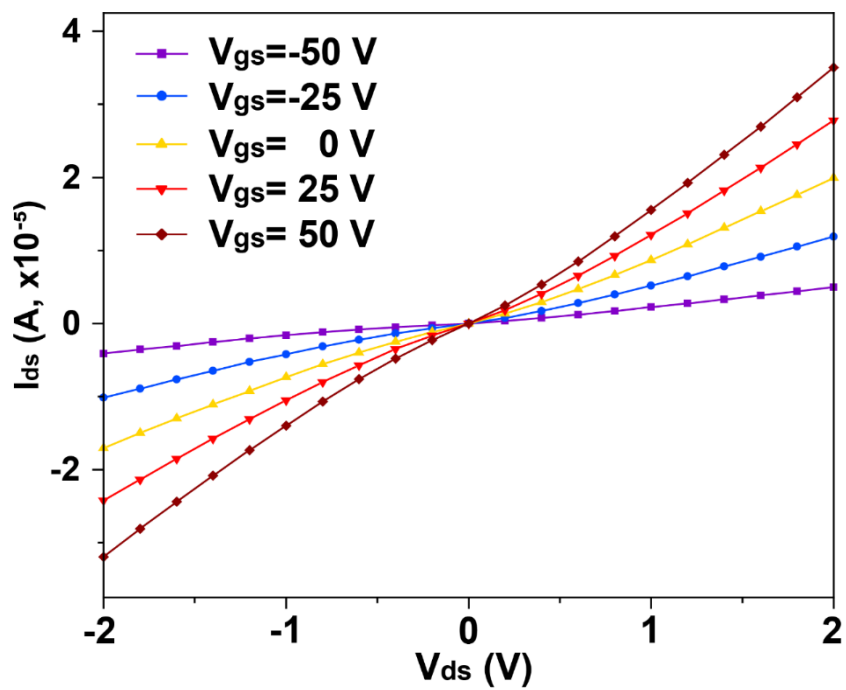

**Fig. S14. The output curves of monolayer pristine MoS$_2$.**

**Fig. S15.**

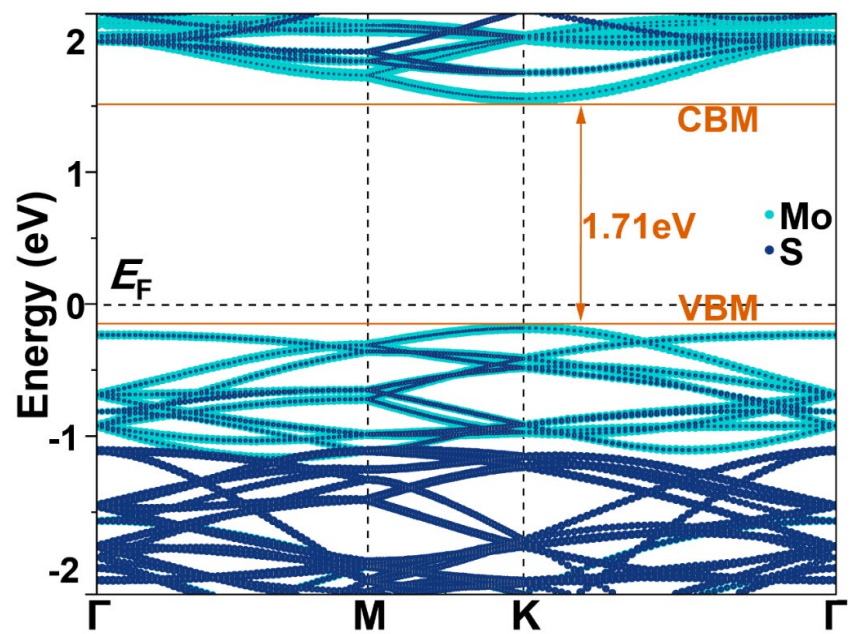

**Fig. S15. The calculated band structure of monolayer pristine MoS$_2$.** The cyan and dark blue dots represent the contribution of Mo and S atoms, respectively.

**Fig. S16.**

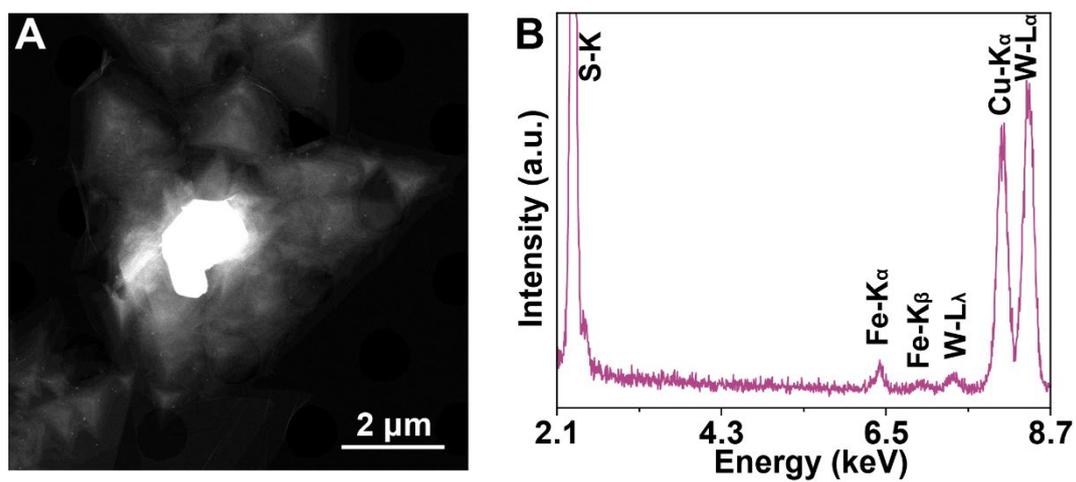

**Fig. S16.** (**A**) The HAADF-STEM image of a Fe-doped WS$_2$ flake. (**B**) EDX spectrum of Fe-doped MoS$_2$ measured from sample in (**A**).

Fig. S17.

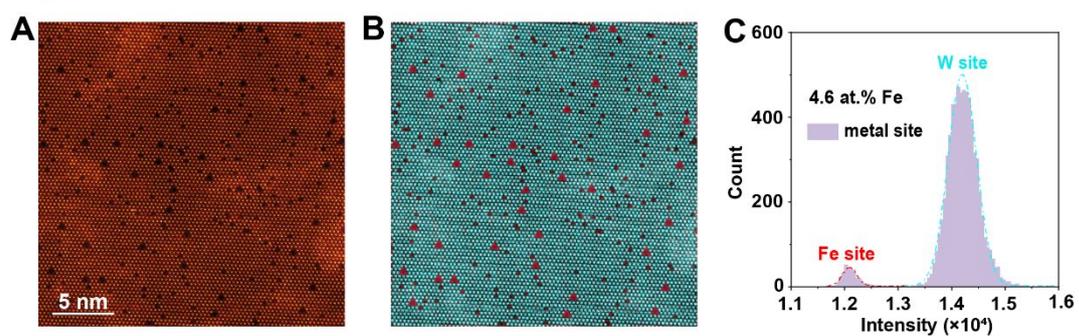

**Fig. S17. Microscopic characterizations of monolayer Fe-doped WS$_2$.** (**A**) Atomic-scale HAADF-STEM images of monolayer Fe-doped WS$_2$. (**B**) The labelled sites of W atoms (cyan circles) and Fe atoms (red circles) for monolayer Fe-doped WS$_2$. (**C**) The statistical charts of atomic intensity for labelled atoms in monolayer Fe-doped WS$_2$.

**Table S1.**

**Table S1**. **Calculated Raman vibrational modes of monolayer pristine MoS$_2$.** Phonon frequencies and irreducible representations of monolayer MoS$_2$ at Γ, M, and K high-symmetry points from our DFT calculations. Groups of the wave vector are also labeled.

| Γ($D_{3h}$) | $\omega(cm^{-1})$ | M($C_{2v}$) | $\omega(cm^{-1})$ | K($C_{3h}$) | $\omega(cm^{-1})$ |
|---|---|---|---|---|---|
| $A_2''(ZO)$ | 453.8 | $A_1$ | 390.7 | $A'$ | 387.9 |
| $A_1'(ZO)$ | 391.6 | $B_1$ | 368.2 | $E''$ | 360.7 |
| $E'(LO)$ | 366.0 | $B_2$ | 349.3 | $E'$ | 352.8 |
| $E'(TO)$ | 366.0 | $A_1$ | 340.1 | $E'$ | 316.0 |